\documentclass[iop]{emulateapj}

\newcommand{\gsim}{\;\lower.6ex\hbox{$\sim$}\kern-7.75pt\raise.65ex\hbox{$>$}\;}
\newcommand{\lsim}{\;\lower.6ex\hbox{$\sim$}\kern-7.75pt\raise.65ex\hbox{$<$}\;}

\begin{document}


\title{The Star Formation History of the very metal poor BCD I~Zw~18 from HST/ACS data\footnote{Based on observations with the NASA/ESA {\it Hubble Space Telescope}, obtained at the 
Space Telescope Science Institute, which is operated  by AURA, Inc.,
for NASA under contract NAS5-26555.}}


\author{F. Annibali \altaffilmark{2}, M. Cignoni \altaffilmark{2,3}, M. Tosi \altaffilmark{2},  R.~P. van der Marel \altaffilmark{4}, 
A. Aloisi \altaffilmark{4}, G. Clementini \altaffilmark{2}, R. Contreras Ramos \altaffilmark{2}, G. Fiorentino \altaffilmark{2}, 
M. Marconi \altaffilmark{5}, I. Musella \altaffilmark{5}.}

\altaffiltext{2}{INAF-Osservatorio Astronomico di Bologna, 
Via Ranzani 1, I-40127 Bologna, Italy; francesca.annibali@oabo.inaf.it}

\altaffiltext{3}{Dipartimento di Astronomia, Universit\`a degli Studi di Bologna, 
Via Ranzani 1, I-40127 Bologna, Italy}

\altaffiltext{4}{Space Telescope Science Institute, 3700 San Martin Drive, 
Baltimore, MD 21218, USA}

\altaffiltext{5}{INAF-Osservatorio Astronomico di Capodimonte, via Moiariello 16, 80131 Napoli, Italy}



\begin{abstract}

We have derived the star formation history (SFH) of the blue compact dwarf galaxy I~Zw~18 through comparison of deep HST/ACS data 
with synthetic color magnitude diagrams.  A statistical analysis was implemented for the identification of the best-fit SFH and relative uncertainties. 
We confirm that I~Zw~18 is not a truly young galaxy, having started forming stars earlier than $\sim$1 Gyr ago, and possibly at epochs as old as a Hubble time. 
In I~Zw~18' s main body we infer a lower limit of  $\approx 2 \times 10^{6} M_{\odot}$ for the mass locked-up in old stars.
I~Zw~18' s main body has been forming stars very actively during the last $\sim$10 Myr, 
with an average star formation rate (SFR) as high as $\approx 1 M_{\odot}/yr$ (or $\approx2 \times10^{-5} M_{\odot} yr^{-1} pc^{-2}$). 
On the other hand, the secondary body  
was much less active at these epochs, in agreement with the absence of significant nebular emission. 
The high current SFR can explain the very blue colors and the high ionized gas content in I~Zw~18,   
resembling primeval galaxies in the early Universe. Detailed chemical evolution models are required to 
quantitatively check whether the SFH from the synthetic CMDs can explain the low measured element abundances, 
or if galactic winds with loss of metals are needed.

\end{abstract}





\section{Introduction} \label{sec_intro}

The blue compact dwarf (BCD) galaxy I~Zw~18 is one
of the most intriguing objects  in the local Universe and has fascinated generations
of astronomers since its discovery \citep{zw}.
With a metallicity between 1/30 and 1/50 Z$_{\odot}$
\citep{ss72,lq79,dk85,duf88,pag92,sk93,kunt94,sl96,garnett97,it98}
it holds the record
of the second lowest metallicity and lowest helium content measured 
in a star-forming galaxy  \citep{izotov09}.
The dynamical mass of I~Zw~18 
measured at a radius of 10''-12'' is $\sim 2-3\times10^8$ M$_{\odot}$  
\citep{petrosian97,vanzee98,lelli12}. 
A large amount of gas,
corresponding to $\sim 70\%$ of the total mass, is detected all around
the system, but only  $10^7 M_{\odot}$ of HI is associated with the 
optical part of the galaxy \citep{lv80,vanzee98}.
The HI associated to the starburst region forms a compact rapidly rotating disk, 
and the steep rise of the rotation curve in the inner parts indicates 
that there is a strong central concentration of mass \citep{lelli12}. 
HI observations have revealed a neutral hydrogen bridge
connecting I~Zw~18 with a faint companion galaxy (the so called C component or secondary body), 
which has also been resolved into stars with HST imaging \citep{dou96,it04,alo07}, and 
a $\sim$13.5 kpc HI tail extending to the south of I~Zw~18 main body \citep{lelli12}. 

I~Zw~18 shows very blue colors, $U-B=-0.88$ and $B-V=-0.03$ \citep{vanzee98}, 
suggesting the presence of a very young stellar population, with a current star formation rate (SFR) much higher than the past mean value 
\citep{ss72}. All these observational
characteristics make I~Zw~18 resemble a primeval galaxy in the nearby Universe.
In fact, soon after its discovery, the question
arose whether I~Zw~18 is so metal poor because a) 
it started forming stars only recently,
so that they haven't had much time to 
pollute metals in the interstellar medium,
or b) because its star formation (SF) activity, 
although occurring over a long period of time, has proceeded at
a rate too low for an efficient chemical enrichment, or c) 
because strong galactic winds have 
removed from the system most of the metals.
The nature of I~Zw~18 has important cosmological implications.
If indeed some BCDs turned out to be young galaxies, their
existence would support the view that SF in low-mass systems has been
inhibited till the present epoch \citep[e.g.,][]{br92}. 
On the other hand, the lack of such primordial
systems would provide strong constraints on chemical evolution and
hydrodynamical models of metal-poor galaxies, e.g., on their SF regime
(continuous or bursting?) or the onset of a galactic wind. 

With the advent of the Hubble Space Telescope (HST),  
it has been possible to resolve  the individual stars in I~Zw~18 and  
thus  to characterize its evolutionary status.
From  HST/WFPC2 data, \cite{ht95} and \cite{dou96} 
argued for a continuous SF over the last 30-50 Myr.
Thanks to a more sophisticated treatment of the same dataset, 
\cite{aloisi99} were 
able to go deeper and to detect asymptotic giant branch (AGB) stars 
with ages of at least several hundreds Myr. 
These results were subsequently confirmed by \cite{ost00} through deep
HST/NICMOS imaging. 
Later, \cite{it04} failed to detect red giant branch (RGB) stars with the Advanced Camera for Surveys (ACS)
on board of HST and concluded that the galaxy is at most 500 Myr old. 
However, both \cite{mom05} and \cite{tosi07}, from independent reanalyses of the same ACS data set, 
suggested that I~Zw~18 should be older than at least 1-2 Gyr, since it does appear to contain also RGB stars.

In order to shed light on the situation, we acquired in 2006 new 
time-series HST/ACS photometry to study Cepheid stars in I~Zw~18 and 
pin down its distance (GO~10586, PI Aloisi). By combining the new data with archival ones, 
we both identified the RGB tip (TRGB) at $I_0=27.27\pm0.14$ mag \citep[$D=18.2\pm1.5$ Mpc, ][]{alo07}, 
and detected for the first time a few Cepheids whose light
curves allowed us to independently derive the distance to $\approx$ 1 Mpc of accuracy 
\citep[D$=19.0\pm0.9$ Mpc, ][]{fiore10,marco10}. 
The detection of RGB stars in I~Zw~18 implies that it has started forming stars at least $\sim$1 Gyr ago, and possibly at  epochs as old as a Hubble time, ruling out the possibility that it is
a truly primordial galaxy formed recently in the local Universe.
In this paper we use synthetic CMDs to reconstruct the entire star formation history (SFH) within the reachable look-back time and to quantify the 
stellar mass formed at old, intermediate-age, and young epochs. 
A qualitative study of the spatial distribution and
evolutionary properties of I~Zw~18' s resolved stars has been
already presented by \cite{contr11}.

\begin{figure}
\epsscale{0.6}
\plotone{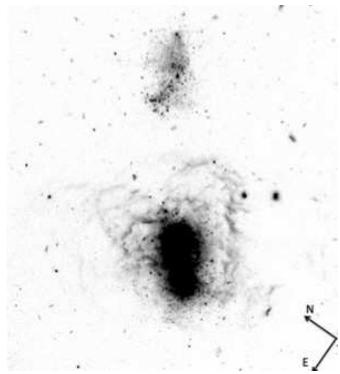}
\caption{F606W (broad V) ACS/WFC image showing I~Zw~18' s main (south-east) and secondary (north west) bodies. 
\label{fig1}}
\end{figure}

\section{Observations and data reduction}  \label{sec_obsred}

Observations and data reduction have been extensively described in 
previous papers \citep{alo07,fiore10,contr11}.  Here we only recall the main aspects. 
The data were collected between October 2005 and January 2006
with the ACS/WFC in 13 different epochs in F606W ($\sim$broad V), and 
12 different epochs in F814W ($\sim$I), for total integration
times of $\approx$ 27,700 s and $\approx$ 26,200 s.
The single exposures per epoch per filter were reprocessed
with the most up-to-date version of the ACS
calibration pipeline (CALACS).
Then, for each filter, we co-added all
the exposures into a single image using the
software package MULTIDRIZZLE \citep{Koe02},
in order to obtain 2 deep images in F606W and F814W, 
respectively. A portion of the F606W deep image showing I~Zw~18' s 
main and secondary bodies is shown in Fig.~\ref{fig1}.
The pixel size of the final resampled 
drizzled images is 0.035 \arcsec (0.7 times the 
original ACS/WFC pixel size). 
We also retrieved and combined archival ACS/WFC data in F555W ($\sim$V) and
F814W (GO program 9400, PI Thuan), taken in 2003 over a period
of 11 days, into two deep master images with total integration times of $\approx$ 43,500 s 
and $\approx$ 24,300 s, respectively.

Photometry was performed with
DAOPHOT \citep{daophot} in the IRAF environment\footnote{IRAF is distributed
by the National Optical Astronomy Observatories, which are operated by
AURA, Inc., under cooperative agreement with the National Science 
Foundation} for both the archival and proprietary datasets.  
The instrumental magnitudes were estimated via a 
PSF-fitting technique. The PSF was created choosing the 
most isolated and clean stars in the vicinity
of I~Zw~18's main and secondary bodies, and was  
modeled with an analytic Moffat function 
plus additive corrections derived from
the residuals of the fit to the PSF stars.
In order to push the photometry as deep as possible, 
stars were detected  in a sum of the V and I images in each dataset,
choosing a detection treshold of 2.4 times the local background level.
This procedure allowed us to recover faint
objects with very blue or red colors. 
Aperture photometry with PHOT, and then
PSF-fitting photometry with the DAOPHOT/ALLSTAR package
\citep{daophot}, were performed separately
on the V and I images at the position 
of the objects detected in the sum image.
The finding procedure, the aperture photometry,
and the PSF-fitting photometry were then 
re-iterated on the subtracted V and I images.
This allowed us to recover additional faint stars
showing up only after the subtraction of their brighter 
neighbours. For each dataset, the HST magnitudes were calibrated into the Johnson-Cousins system using the transformations in \cite{sir05}.
The CMD presented in \cite{alo07} was obtained 
by cross-correlating the photometry in the two datasets 
requiring good matches in position (within one pixel) and magnitude (within 3.5$\sigma$), and a Daophot sharpness 
parameter $\le0.5$.  
 The final V and I magnitudes were obtained averaging the magnitudes in the two datasets.  
In this paper, we use this matched catalog with $\approx$2000 sources to derive the SFH of I~Zw~18.

\section{Color-Magnitude diagrams} \label{sec_cmd}

\begin{figure*}
\epsscale{1}
\plotone{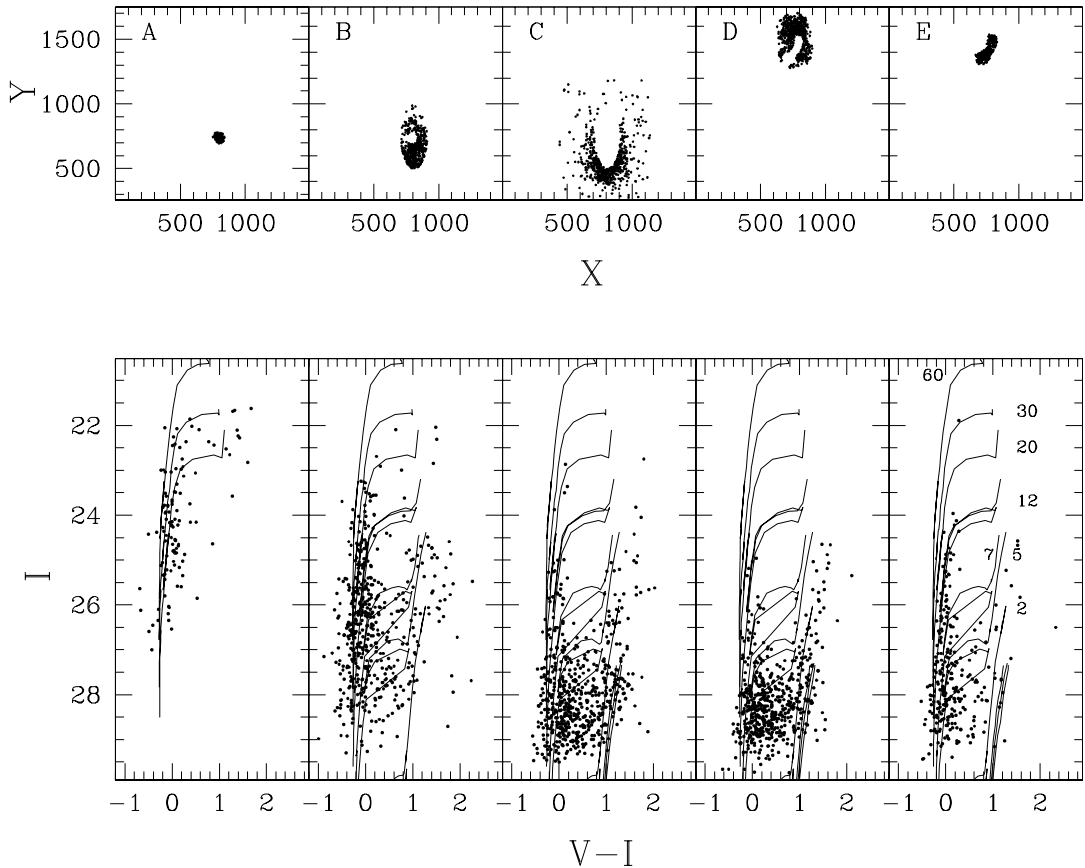}
\caption{Upper panels: spatial distribution of the stars in regions A, B, C, D and E selected in I~Zw~18.  
Bottom panels: corresponding CMDs. Superimposed are the Padova~94 stellar tracks for the masses: 60, 30, 20, 12, 7, 5, 2, 1, 0.8, and 0.6 $M_{\odot}$.
For regions A and B, the smallest plotted masses are 20 and 2  $M_{\odot}$, respectively. 
\label{fig2}}
\end{figure*}

To account for the highly variable crowding within I~Zw~18, we selected five regions (A, B, C in the main body, D and E in the secondary body). The corresponding  I, V$-$I color-magnitude diagrams (CMDs)  are shown in Fig.~\ref{fig2}. In the main body, region A is the most crowded one, and 
is embedded into a shell of ionized gas, while region C, at the outskirts, is the least crowded one. 
In the secondary body, region E is more crowded than region D. 
We detect 111, 481, 590, and 548, and 265 sources in regions A, B, C, D, and E, over an area of $\approx$0.06, 0.66, 4.83, 0.17, and 0.70 kpc$^2$, respectively.  However, in region C the majority 
 of stars are concentrated into a smaller area of 0.76 kpc$^2$, and thus we will refer to this value hereafter  to compute 
 the specific star formation rate (see Table~\ref{tab_sfr}).
The 8.7-day period classical Cepheid of \cite{fiore10} is located in region C, the other two Ultra long-period Cepheids 
 \citep[P$\sim$130 d, see also][]{fiore12} are found in regions C and A, while the two long period variables 
 (P$>$100 d) are in regions B and C.

 On the CMDs in Fig.~\ref{fig2}, we have overplotted the Padova~94 \citep{pad94} stellar tracks for different masses to show the mass ranges (and thus the look-back times) sampled in the different regions.  
In region A, a concentration of bright stars between the 20 and 30 $M_{\odot}$ tracks, with ages 7-10 Myr, and a few stars with mass as low as $\sim$ 12 $M_{\odot}$ ($\sim$20 Myr old) are detected. Lower mass stars are lost because of the severe photometric incompleteness due 
to the extreme crowding. In region B, the look-back time sampled is longer than in region A, with stellar masses as low as 2 $M_{\odot}$ ($\sim$ 1 Gyr old) present on the CMD. 
However, only in the least crowded region C  red giant branch (RGB) stars  with masses $<  2 M_{\odot}$ and 
ages  $>$ 1 Gyr are detected. This also holds for region D in the secondary body, while in the more crowded region E only 
a few potential RGB stars are observed. From the comparison of the CMDs with the theoretical tracks, we also notice the 
clear presence of an age gradient moving from the most crowded regions to the least crowded ones. In fact, the bright part of the blue plume 
 (I$<$26 mag) is significantly more populated in regions A, B and E than in regions C and D, and reaches brighter magnitudes. This implies 
 more massive stars and younger star formation. For a quantitative derivation of the star formation history (SFH) in the different regions 
 we refer to Section~\ref{sec_sfh}. 

\section{Artificial star experiments} \label{sec_artificial}

To evaluate the role of incompleteness and blending in the data, 
we performed artificial star experiments on the original frames,
following the same procedure described in \cite{anni07}.
These tests serve to probe observational effects 
associated with the data reduction process, such as the accuracy 
of the photometric measurements, the crowding conditions, and 
the ability of the PSF-fitting procedure in resolving partially 
overlapped sources.
We performed the tests for the individual F555W, F606W, and for the 
two F814W frames, according to the following procedure.
We divided the frames into grids of cells of chosen width
(50 pixels) and randomly added one artificial star per cell at each run.
This procedure prevents the artificial stars
to interfere with each other, and avoids to bias the experiments
toward an artificial crowding not present in the original
frames. The position of the grid is randomly changed at each
run, and after a large number of experiments the stars are uniformly distributed over the frame. 
In each filter, we assign to the artificial star a random input magnitude 
between m1 and m2, with m1 $\approx$ 3 mag brighter 
than the brightest star in the CMD, and m2 $\approx$ 3 mag fainter 
than the faintest star in the CMD. At
each run, the frame is re-reduced following the same
procedure as for the real data. 
To account for the fact that in the real data 
the source detection was performed on a sum of the V$+$I frames, we adopted a smaller    
treshold than the value of 2.4 adopted for the real data. 
The new treshold is derived performing tests on the original frames, and 
is such that the number of detections in a single frame equals 
the number of detections in the sum of the V$+$I frames 
with a treshold of 2.4.
The output photometric catalog
is cross-correlated with a sum of the original photometric catalog 
of real stars and the list of the artificial stars added into
the frame. This prevents cross-correlation of artificial stars in
the input list with real stars recovered in the output photometric
catalog. We simulated $\sim$500,000 stars for each filter.
Stars with input-output magnitude $>$ 0.75 
were considered lost, because such a difference  
implies that they fell on a real star of their same luminosity 
or brighter. 
For each magnitude bin, the completeness of our photometry was 
computed as the ratio of the number of recovered artificial stars
over the number of added ones. 

\begin{figure*}
\epsscale{1}
\plotone{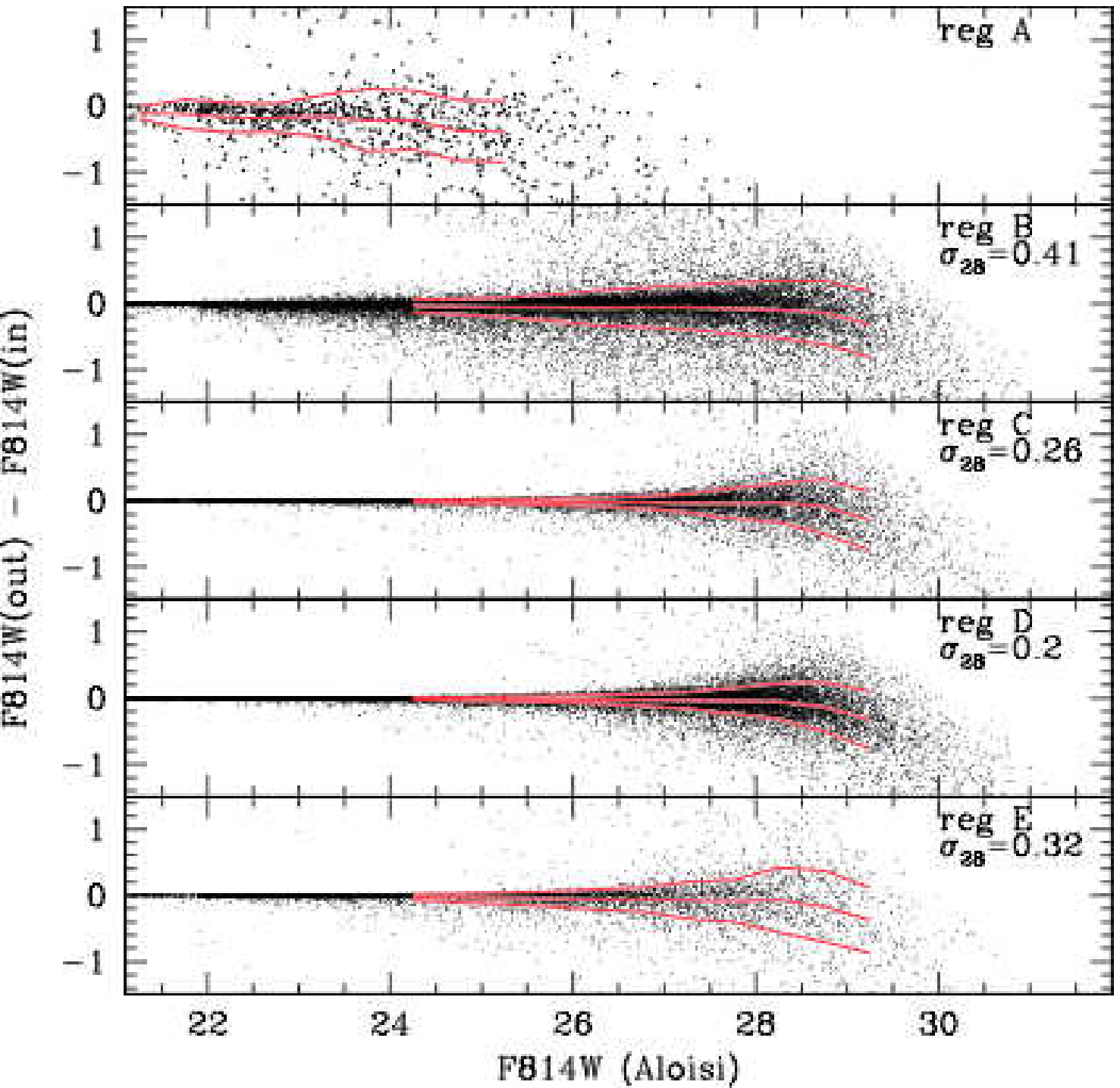}
\caption{Output minus input magnitude versus input magnitude in F814W for the artificial
stars simulated in the Aloisi dataset.  The distribution is shown for the different regions 
(A, B, C, D, E) selected in the I~Zw~18' s  field. 
The solid curves indicate the mean of the $\Delta$mag distribution (central line) and the 
$\pm 1 \sigma$ standard deviations. For regions B, C, D and E, we give the  
$\sigma$ value at F814W$=$28.
\label{fig3b}}
\end{figure*}

\begin{figure}
\epsscale{1}
\plotone{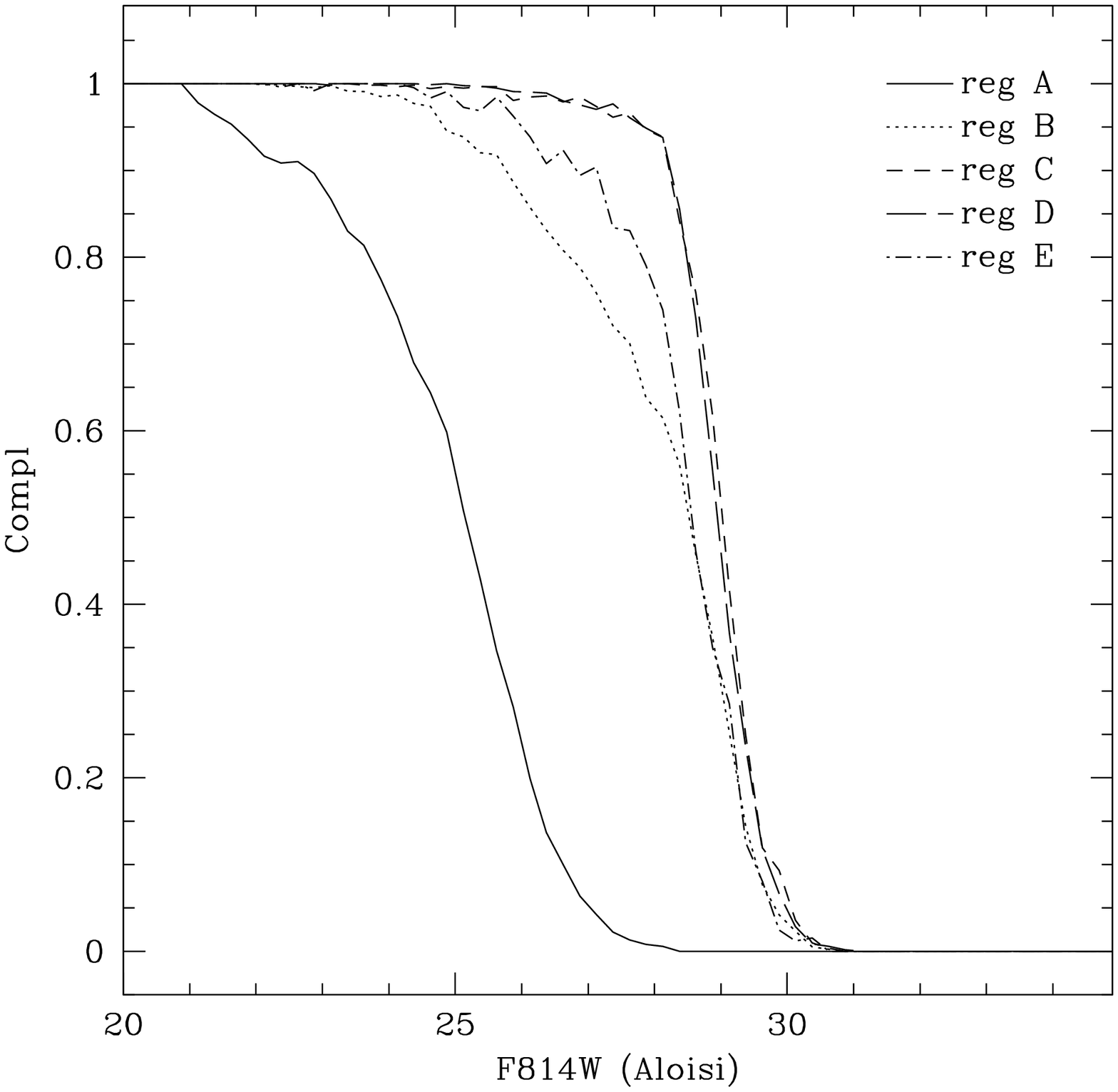}
\caption{Completeness factor (1$=$100\% complete) as a function of the F814W magnitude 
for the Aloisi dataset. The different lines correspond to the different 
regions A, B, C, D, and E selected in I~Zw~18, as indicated by the labels.   
\label{fig3a}}
\end{figure}

As an example, we show in Fig.~\ref{fig3b} the distribution of the artificial star 
output minus input magnitude ($\Delta$m) in F814W for the Aloisi dataset. We show the different 
distributions relative to regions A, B, C, D and E selected in I~Zw~18, in order to 
characterize the photometric error as a function of magnitude and crowding. 
The systematic deviation from 0 of the mean $\Delta$m 
is due to the increasing effect
of blending at fainter magnitudes: faint stars tend to be systematically recovered with brighter magnitudes 
because they happen to overlap with objects of comparable or brighter magnitude. 
The photometric errors systematically decrease from the most crowded region A to the least crowded regions C and D.
At the same time, the completeness increases, as shown in Fig.~\ref{fig3a}.

\section{Star Formation History} \label{sec_sfh}

\subsection{Methodology}

We derived the star formation history (SFH) of I~Zw~18 through comparison of the observed CMDs with synthetic ones. 
Our procedure consists of two separate steps as described in the following subsections: 1) the definition of the so called {\it basis functions} from the combination of the adopted stellar evolution models with the photometric 
 properties of the examined region, and 2) the statistical analysis for the identification of the best SFH and relative uncertainties. In order to better characterize the uncertainties intrinsic in the statistical analysis, we applied two completely independent procedures: the one described by \cite{gro12} (hereafter, the Baltimore procedure) and that described by  \cite{cigno10} (hereafter, the Bologna procedure). The comparison between the two methods provides important information on which features of the SFH are robust and which are instead artifacts due to the adopted minimization approach. 
On the other hand, quantifying the systematics due to the choice of a particular set of stellar models goes beyond the scope  of this paper. 
 
\subsubsection{Basis Functions}

First, we created a grid of synthetic CMDs for episodes with $\log(age)$ from 6 to 10, and duration $\delta log(age)=0.25$ (hereafter, {\it{basis functions}}). The choice of logarithmic 
steps allows us to account for the increasingly lower time resolution in the SFH at larger look-back time. 
The basis functions are created following the approach initially described in \cite{tosi91}, and 
adopting the last version of the code by \cite{ange05}.  
In brief, the synthetic CMDs are produced via Monte Carlo extractions of (mass, age) pairs for 
an assumed Initial Mass Function (IMF), adopting a constant SF from the starting to the ending epoch of each basis function. 
Here we adopt a Salpeter' s IMF \citep{salp55} and zero binary fraction.   
Each star is placed in the theoretical ($\log L/L_{\odot}$, $\log T_{eff}$) plane via interpolation on a chosen set of stellar evolutionary tracks 
\citep[here, the Padova tracks with $Z=0.0004$, ][]{pad94}. Luminosity and effective temperature are transformed into 
the ACS Vegamag F555W, F606W, and F814W filters using the \cite{ol00} code. Absolute magnitudes 
are then converted into apparent ones applying a foreground reddening of $E(B-V)=0.032$ from the NED and 
a distance modulus of $(m-M_0)=31.3$ mag derived by \cite{alo07}. 
At this point we apply a completeness test in order to determine 
whether to retain or to reject the synthetic star, based on the results 
of the artificial star tests performed in all four images (F606W and F814W for the Aloisi dataset, and 
F555W and F814W for the Thuan dataset, see Section~\ref{sec_artificial}). 
Since we are using the CMD obtained from the cross-correlation of four photometric catalogs   
(see Section~\ref{sec_cmd}), we require in the simulations that the synthetic stars pass the test in 
all four photometric bands. 
Photometric errors are assigned on the 
basis of the distribution of the output-minus-input magnitudes 
of the artificial stars (Fig.~\ref{fig3b}). These errors take into account the various 
instrumental and observational effects, as well as systematic 
uncertainties due to crowding (i.e., blend of fainter objects into 
an apparent brighter one). Four magnitudes  (F555W, F606W, plus 2 in F814W) 
are associated to each synthetic star. 
The extraction of (mass, age) pairs is stopped when the 
number of stars populating the synthetic CMD equals a certain fixed number. 
This number was chosen to be the same for all the basis functions and to guarantee 
that all the phases potentially sampled in 
the synthetic CMD (given the completeness function) turn out to be well populated.
For a direct comparison with the observed CMDs in Section~\ref{sec_cmd}, the ACS Vegamag 
magnitudes are transformed into the Johnson-Cousins system and are averaged following the same 
procedure as for the real data. The same basis functions are implemented into the Baltimore and the 
Bologna procedures.  

\subsubsection{Statistical analysis with the Baltimore approach}

The SFH was derived through a statistical approach using the code SFHMATRIX  
developed by R. van der Marel in Baltimore and described in 
\citep{gro12}. 
In brief, the SFH is inferred finding the weighted combination of basis functions that best
reproduce the observed CMD in a $\chi^2$ sense. To maximize the likelihood, the code considers 
the density of points on the observed/synthetic CMD, i.e. the Hess diagram, 
rather than the CMD itself. The best fit is found by solving a non-negative
least-squares matrix problem.
In SFHMATRIX, the errors on the SFH are calculated 
creating many realizations of pseudo data sets with properties similar to the real data, and analyzing 
these pseudo data in the same way as for the real data. 
Then, the rms scatter in the SFH for a given age is the error bar. 
We create the pseudo-data by drawing many Monte-Carlo
realizations from the best-fit SFH that was inferred from the data.

\subsubsection{Statistical analysis with the Bologna approach}

We also derived the SFH using the Bologna code 
\citep[for details and application see][]{cignoni11,cignoni12},
which has many features in common with the Baltimore code, but also
incorporates some differences in the search of the best synthetic
CMD. Similarly to the Baltimore algorithm, the SFH is parametrized as a
linear combination of quasi simple stellar populations with variable
duration (the basis functions), generated from a specific set of stellar 
models with given metallicity.  
The code searches for the linear combination of the basis functions 
that best matches the observed CMD. The comparison between the observed and the 
synthetic CMDs is performed through a $\chi^2$ minimization in strategic regions 
of the CMD. As the Baltimore code, the Bologna code considers the Hess diagram 
rather than the CMD itself to perform the minimization. The sizes and distribution of the bins within the color-magnitude plane can be variable, and are chosen considering some aspects such as   
the particular CMD shape, the number statistics, the presence of phases where the stellar 
evolution models are more uncertain, and so on. 
Differently from the Baltimore code, the Bologna code employs 
for the minimization a downhill simplex ``amoeba'' algorithm which is re-started
from thousands of initial random positions. A bootstrap method is 
used to assess the statistical uncertainty around the best-fit 
parameters. The minimization process is repeated for each bootstrapped data
set, and the final error bars represent the mean deviation using 1000 bootstraps.

\subsection{CMD fit}

\subsubsection{Region A}

\begin{figure}
\epsscale{1}
\includegraphics[width=9cm]{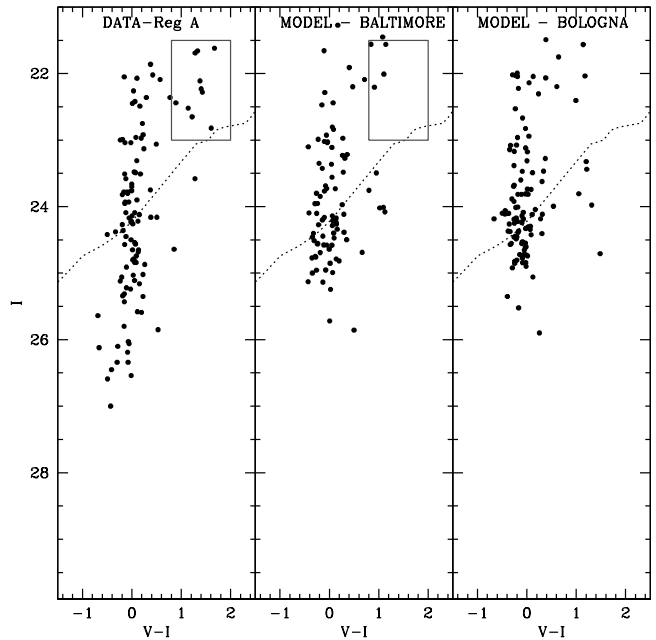}
\caption{From left to right: observed CMD for region A, best-fit synthetic CMD with the Baltimore code, best-fit CMD with the Bologna code.   
The best-fit CMDs were obtained considering only the region above 
the 60\% completeness line, indicated by the dotted curve. 
The box includes red supergiants with age 7-10 Myr.     
\label{fig4}}
\end{figure}

Fig.~\ref{fig4} shows the comparison between the observed CMD selected in the most crowded region A (left) and  
two synthetic CMD realizations drawn from the best-fit SFH with the Baltimore code (middle) and with the Bologna code (right). In both cases, when searching for the best-fit solution, we excluded the region below the 60\% completeness limit from the $\chi^2$ minimization. 
The reason is that crowding is extreme in region A, implying that the derived completeness curve may become highly uncertain below this limit. 
When running the Baltimore code, we adopted a constant pixel size of 0.5 
in both color and magnitude. Since the stellar tracks fail to reach the reddest supergiants in the observed CMD, we decided 
to treat the box at $0.8<V-I<2$, $21.5<I<23$ as a single pixel. 
On the other hand, in the Bologna procedure, the CMD was 
divided into a regular grid of color-magnitude cells with $\Delta(V-I)=0.25$ mag and $\Delta I=1$ mag. 
The choice of the pixel size, which is driven by the need of minimizing statistical fluctuations 
without loosing the information in the CMD, remains somewhat subjective. Thus, it is interesting to see 
how the results change according to the different choices of the grid.  
Both the Baltimore and the Bologna codes provided reduced $\chi^2$ values less than 1 for the best-fit solutions. 
The absolute $\chi^2$ of the fit gives a measure of how well the data are reproduced,
but it is often difficult to interpret in practice. For example, the
effective number of degrees of freedom is difficult to estimate when
many Hess diagram pixels are empty (such as in region A), and do not
effectively contribute to constraining the SFH. We therefore prefer to
assess the quality of fit in this paper through other metrics,
including data-model comparisons of luminosity functions. 

\begin{figure}
\epsscale{1}
\includegraphics[width=9cm]{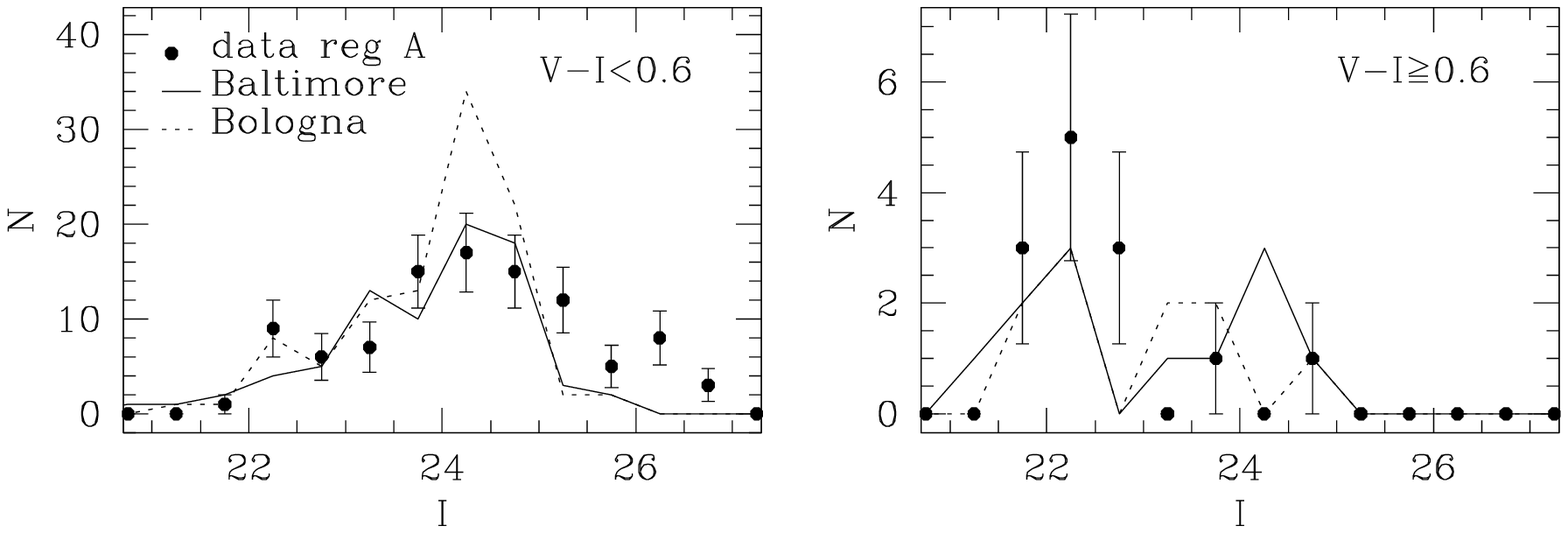}
\caption{Luminosity functions in the blue (left) and in the red (right) for the observed CMD in region A (points) and the best-fit CMDs 
obtained with the Baltimore (solid line) and the Bologna (dotted line) codes.     
\label{fig4b}}
\end{figure}

\begin{figure}
\epsscale{1}
\plotone{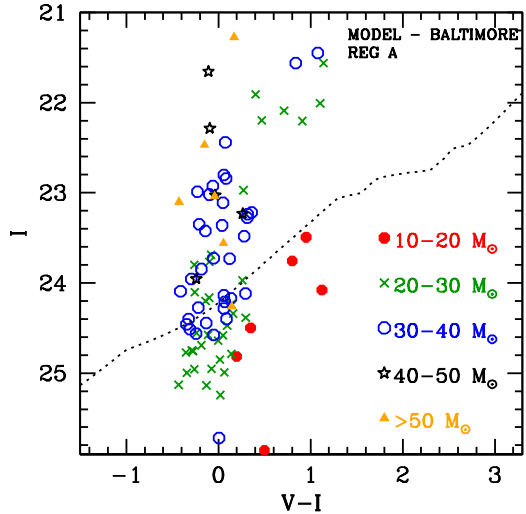}
\caption{Best-fit CMD with the Baltimore code for region A. Different symbols correspond to different masses, as indicated by the labels. 
\label{fig4c}}
\end{figure}

\begin{figure}
\epsscale{1}
\includegraphics[width=9cm]{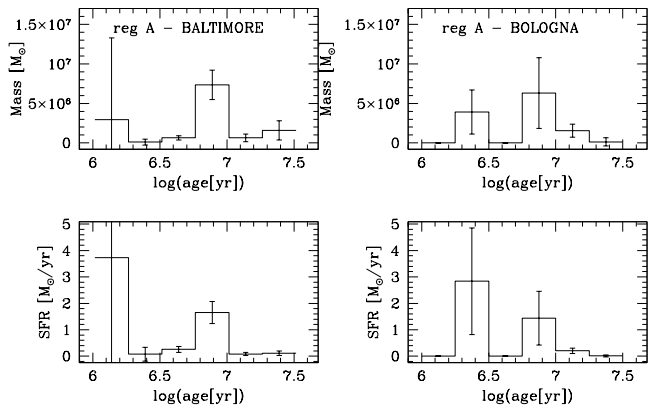}
\caption{Best-fit SFH for region A with the Baltimore code (left panels) and with the Bologna code (right panels). From top to bottom: 
stellar mass formed and SFR in the different age bins. 
\label{fig5}}
\end{figure}

Fig.~\ref{fig4} shows that both the Baltimore and the Bologna solutions well reproduce the overall 
morphology of the observed CMD; however, an inspection of the luminosity functions (LFs) in Fig.~\ref{fig4b}  
reveals some discrepancies between observations and simulations. More specifically, the Baltimore solution tends to underpredict 
the blue ($V-I<0.6$) counts at $22<I<22.5$, and the red ($V-I\ge0.6$) counts at $22<I<23$; these stars are blue and red supergiants 
with masses higher than $\sim$ 15 $M_{\odot}$ and ages less than $\sim$10 Myr in their post-main sequence (post-MS) phase (we will ignore the shortage of simulated stars 
at $I>25.5$ mag, since this region was excluded from the
fit and the completeness correction may be uncertain). 
Also the Bologna solution tends to underpredict the number of red supergiants, but provides a better match to the bright blue supergiants at 
$22<I<22.5$. However, this occurs at the expenses of an excess of fainter blue counts at  $24<I<25$. 
Concluding, we notice that both solutions  tend to under reproduce the correct ratio of brighter over fainter massive post-MS stars.  
As shown in Fig.~\ref{fig4c}, where we have coded the mass in the simulated CMD, this problem can not be solved with a flatter 
IMF, because  the same masses populate at the same time both bright and faint regions in the CMD. 
 Possible explanations are: 
a) the completeness behaviour as a function of magnitude is not properly taken into account; indeed the severe crowding in region A may prevent 
a robust determination of the completeness function through artificial star tests; 
b) some of the brightest stars in the CMD are in fact blends of two or more stars or unresolved clusters rather than individual stars; 
however, as already discussed in \cite{contr11}, the colors of the reddest ($V-I\gsim1$) bright objects in I~Zw~18 are incompatible with the cluster hypothesis; we can also exclude that the large number of bright red stars observed in the CMD is due to reddening within I~Zw~18; in fact, the largest reddening 
value estimated by \cite{cannon02}  in the NW component from the H$\alpha$/H$\beta$ flux ration amounts to just $E(V-I)\sim0.1$; 
c) the evolutionary timescales in the brightest post-MS phase for massive stars are underestimated. The same discrepancy persists 
if we run some  simulations with the new Padova PARSEC stellar models \citep[][and Bressan et al., private communication]{bressan13}
created with a thorough revision of the major input physics, a new treatment of the opacities consistent with the chemical composition,  
and a new treatment of the mass loss. It is possible that models including additional input physics, such as  stellar rotation 
\citep{geneva}, could provide a better match to the brightest objects in the CMD, but these models are not available 
at the moment for a large set of masses at I~Zw~18's metallicity.

The best-fit SFHs obtained with the two different procedures are shown in Fig.~\ref{fig5}. 
For both the Baltimore and the Bologna code, we show the stellar mass formed and the SFR in the different 
age intervals. The results are also summarized in  
Tables~\ref{tab_sfr} and \ref{tab_mass}.  
The values are normalized to a Salpeter IMF in the mass interval 0.1 to 120 $M_{\odot}$. 
The symbol $>$ implies that the value is a lower limit. This is because the look-back time may 
not be large enough to probe the entire age-range in some regions.
The look-back time in region A is $\sim$20 Myr. Both solutions show a strong SF event occurred 6-10 Myr ago with a rate as high $\approx 1.5 M_{\odot}/yr$, comparable to the high rate derived in NGC~1569 \citep{greggio98}. 
A mass of $\approx 6 \times 10^6 M_{\odot}$, similar to that of the largest globular clusters ever discovered 
(e.g. $\omega$ Cen in the Milky Way and G1 in M~31) was formed during this episode. If the SFR is divided by the area sampled by Region A, we obtain $\sim2.5 \times10^{-5} M_{\odot} yr^{-1} pc^{-2}$, e.g. just a few factors lower than the rate derived in the strong star forming region 30~Doradus in the LMC. 
From our best-fit SFH
we can not exclude the presence of an even more recent burst, or ongoing SF.

\subsubsection{Region B}

\begin{figure}
\epsscale{1}
\includegraphics[width=9cm]{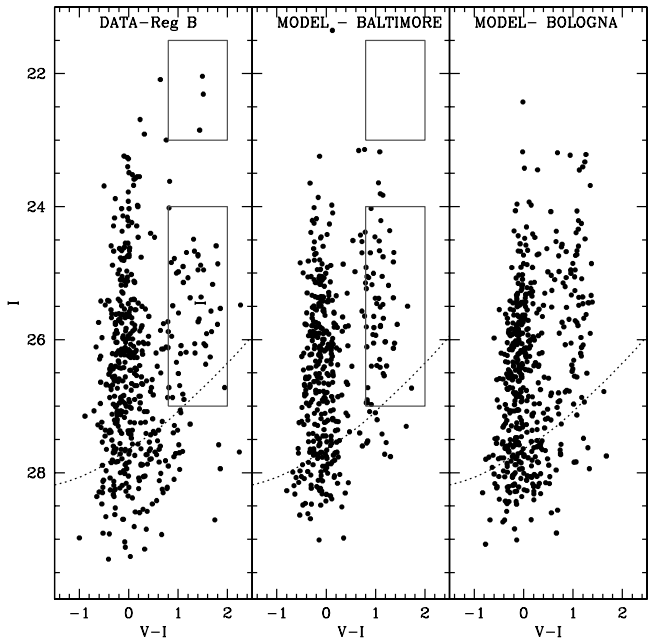}
\caption{
From left to right: observed CMD for region B, best-fit synthetic CMD with the Baltimore code, best-fit CMD with the Bologna code.   
The best-fit CMDs were obtained considering only the region above 
the 40\% completeness line, indicated by the dotted curve. 
The upper box includes red supergiants with age 7-10 Myr. The lower box is where TP-AGB stars are located. 
\label{fig6}}
\end{figure}

The comparison between the observed CMD in region B and two synthetic CMD realizations  
from the best-fit SFH with the Baltimore and the Bologna codes is shown in Fig.~\ref{fig6}.
We excluded from the fit the area below the dotted line where the completeness is below 40\%. 
Given the higher density of stars on the CMD than in region A, we adopted here a smaller pixel size of 0.25 
in both color and magnitude with the Baltimore code, while we chose a pixel size of 0.25 in color and 0.5 in magnitude to 
run the Bologna code. 
When running the Baltimore code, we treated the box at $0.8<V-I<2$, $21.5<I<23$ as a single pixel 
(as for region A), and excluded the  box at  $0.8<V-I<2$, $24<I<27$, where TP-AGB stars are located, 
from the fit. The reason for this is that the TP-AGB phase is not included in the Padova~94 
tracks that we use in this paper. 
Although more recent Padova models \citep[see ][]{girardi10} provide a much 
improved treatment of the TP-AGB stars, this is still a complex and highly 
uncertain phase of stellar evolution. Moreover, a significant but 
unknown fraction of TP-AGB stars may be obscured by dust 
shells \citep{boyer09} making it difficult to use such stars to constrain the SFH. 
On the other hand, the Bologna code was run considering the entire CMD down to the 
40\% completeness limit. We purposely adopted a different approach for the TP-AGB region in
the two codes to be able to assess the impact of this region on the final inferred SFH. This is valid for all the other regions as well. 

\begin{figure}
\epsscale{1}
\includegraphics[width=9cm]{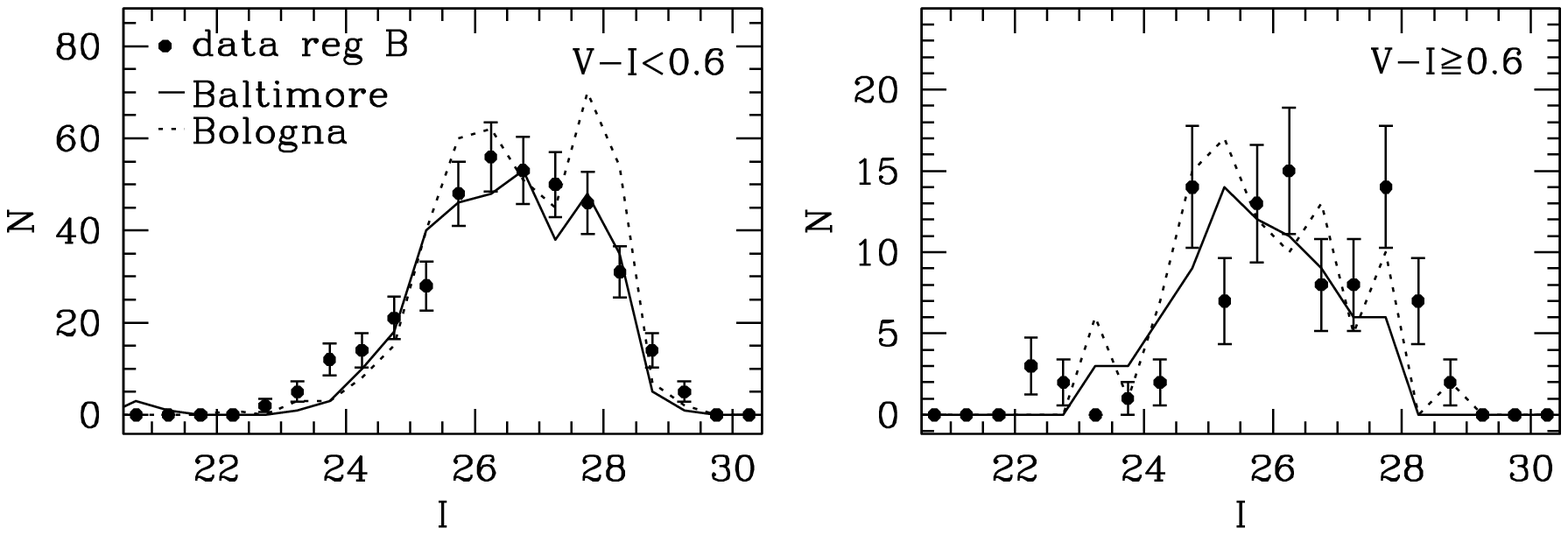}
\caption{Luminosity functions in the blue (left) and in the red (right) for the observed CMD in region B (points) and the best-fit CMDs 
obtained with the Baltimore (solid line) and the Bologna (dotted line) odes.     
\label{fig6b}}
\end{figure}

\begin{figure}
\epsscale{1}
\includegraphics[width=9cm]{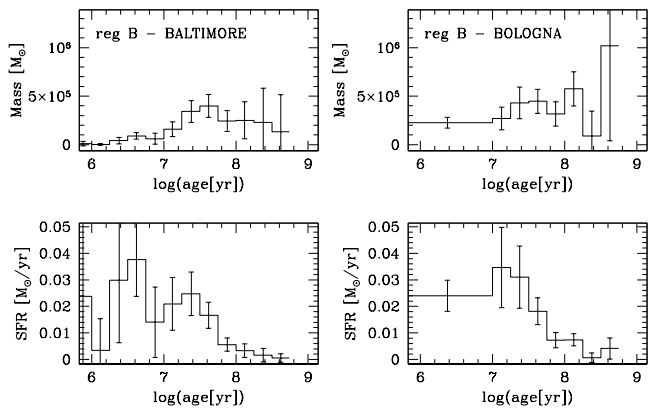}
\caption{Best-fit SFH for region B with the Baltimore code (left panels) and with the Bologna code (right panels). From top to bottom: 
stellar mass formed and SFR in the different age bins. 
\label{fig7}}
\end{figure}

As shown in Fig.~\ref{fig6}, both solutions reproduce quite well the overall morphology of the observed CMD.
The agreement is reasonable also when comparing the observed and  simulated LFs in Fig.~\ref{fig6b}. 
The main discrepancies that we notice are the following: 
1) both solutions tend to under-reproduce the number of bright  blue stars at $I\lesssim24$ mag and bright red stars 
$22\lesssim I \lesssim23$ (we noticed an analogous problem in region A); 
2) the Baltimore solution tends to under reproduce the faint red counts at $I>27.5$ mag, while the Bologna solution over predicts the blue counts at $I>27.5$ mag; 
however, we recall that these regions fall below the 40\% completeness limit and were excluded from the fit when searching for the best-fit solution. 

The best-fit SFHs obtained with the two different procedures are shown in Fig.~\ref{fig7}. 
In both solutions, the SFR decreases with increasing look-back time ($\sim$300 Myr is the maximum look-back time reached in this region);
the highest peak of SF occurred at a rate of $\approx 0.04 M_{\odot}/yr$  during the last 10-20 Myr (depending on the solution, see 
Table~\ref{tab_sfr}).
The total mass formed in region B over the last $\sim$300 Myr is $\approx 2-3\times 10^6 M_{\odot}$, 
depending on the solution (see Table~\ref{tab_mass}).

\subsubsection{Region C}

\begin{figure}
\epsscale{1}
\includegraphics[width=9cm]{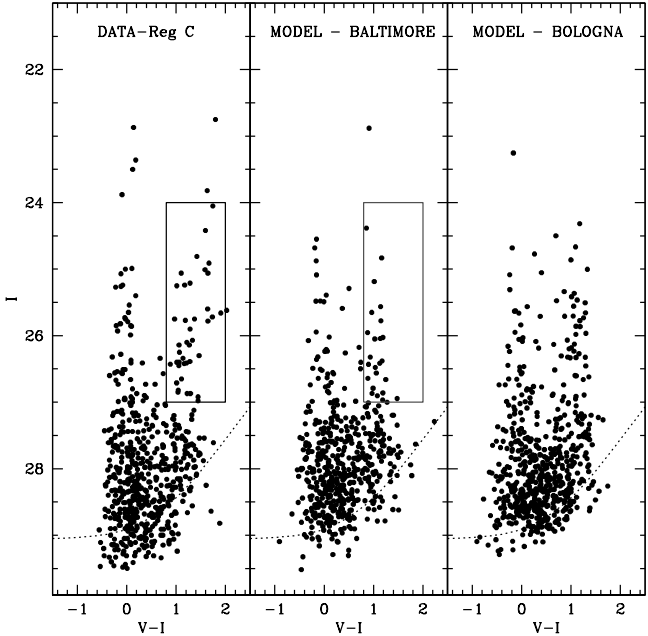}
\caption{
From left to right: observed CMD for region C, best-fit synthetic CMD with the Baltimore code, best-fit CMD with the Bologna code.   
The best-fit CMDs were obtained considering only the region above 
the 20\% completeness line, indicated by the dotted curve. The box is where TP-AGB stars are located. 
\label{fig8}}
\end{figure}

\begin{figure}
\epsscale{1}
\includegraphics[width=9cm]{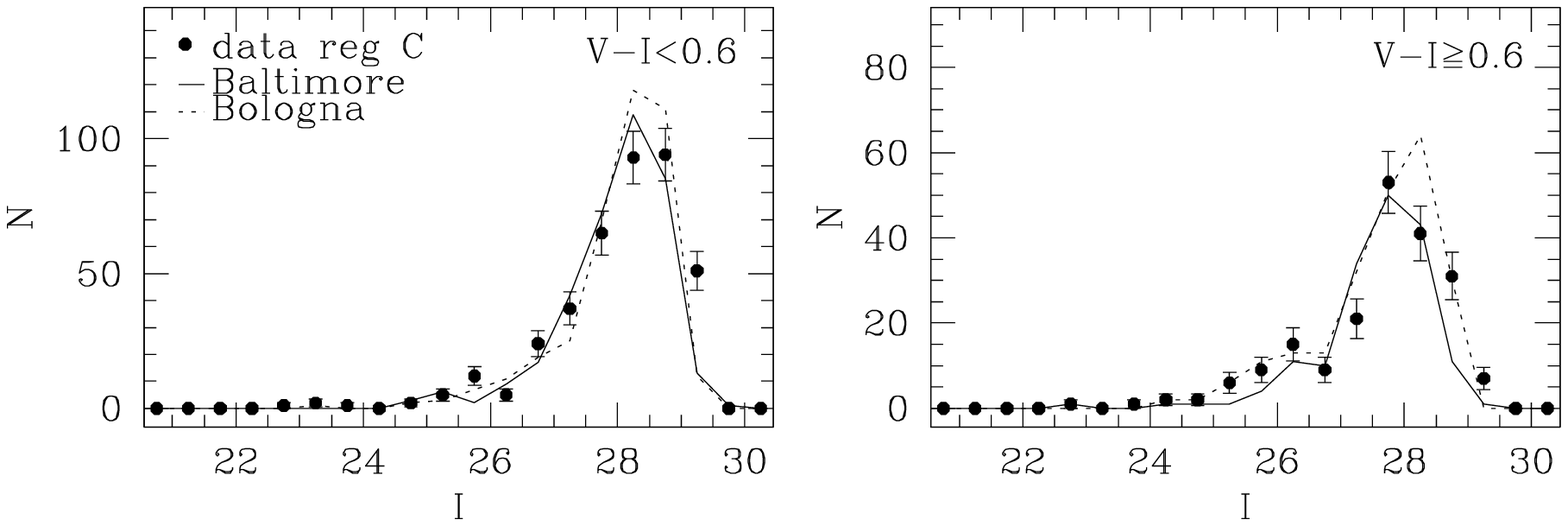}
\caption{Luminosity functions in the blue (left) and in the red (right) for the observed CMD in region C (points) and the best-fit CMDs 
obtained with the Baltimore (solid line) and the Bologna (dotted line) odes.     
\label{fig8b}}
\end{figure}

As shown in Fig.~\ref{fig2}, the CMD of region C is deep enough to reach RGB stars and thus to sample the SF at epochs older than $\sim$1 Gyr. 
The comparison between the observed CMD and two synthetic CMD realizations  
from the best-fit SFH with the Baltimore and the Bologna codes is shown in Fig.~\ref{fig8}.
Here we were less conservative than in regions A and B, and performed  
the  $\chi^2$ minimization using the CMD down to the 20\% completeness limit, in order to reach faint stars 
in the RGB phase (I$\gsim$27.5 mag).  
When running the Baltimore code, we adopted a pixel size of 0.25 in magnitude and color and, as for region B, we excluded from the fit the box at $0.8<V-I<2$, $24<I<27$ where TP-AGB stars are located. On the other hand, the Bologna code was run adopting a pixel size of 0.5$\times$0.5 
and 0.25$\times$0.25 in the magnitude ranges $22<I<25$ and $25<I<29$, respectively, and without the exclusion of CMD regions 
down to the 20\% completeness limit. Both solutions provide a good agreement with the overall CMD morphology in Fig.~\ref{fig8} and with the LFs in Fig.~\ref{fig8b}.  
The Bologna solution tends to exceed the red counts at $28<I<28.5$, while the Baltimore solution tends to under predict the red counts at the faintest 
($I>28.5$) magnitudes;  we recall that this region, where the completeness is very uncertain, was excluded from the fit. 
In general, both solutions provide a good agreement with 
the observed counts at the RGB tip ($I\sim27.5$ mag); here the completeness is $\approx$70\%, 
high enough to allow a reliable derivation of the old ($>$ 1 Gyr) SFH. 

\begin{figure}
\epsscale{1}
\includegraphics[width=9cm]{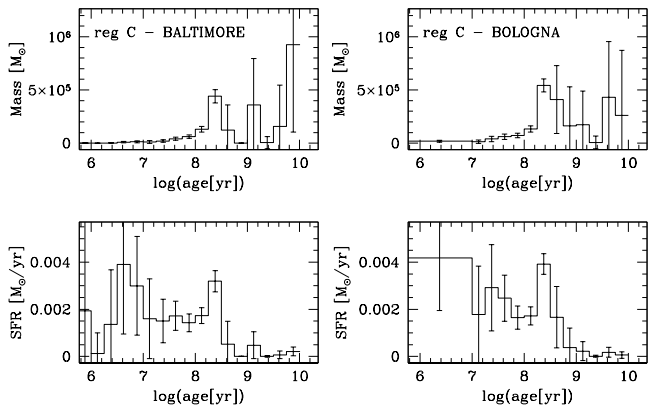}
\caption{
Best-fit SFH for region C with the Baltimore code (left panels) and with the Bologna code (right panels). From top to bottom: 
stellar mass formed and SFR in the different age bins. 
\label{fig9}}
\end{figure}

The best-fit SFHs obtained with the Baltimore and the Bologna procedures are shown in Fig.~\ref{fig9}. 
According to both solutions, the average SFR over the last 100 Myr was $\approx$10 
times as large as the average rate over the 1$-$10 Gyr interval (see Table~\ref{tab_sfr} for details).
However, the lack of resolution at epochs older than 1 Gyr (at a metallicity of $Z=0.0004$ the RGB color is highly degenerate with age and our photometric errors are large) does not allow us to establish if short bursts at a much higher rate occurred in I~Zw~18. 
Indeed, despite the fact that we provide the SFR in four different 
age intervals from $\log(age)=$9 to 10, the only meaningful quantity at these ages is the total stellar mass formed, i.e 
$\approx (0.9 - 1.4)\times10^6 M_{\odot}$, depending on the solution (see Table~\ref{tab_mass}). This is also evident from the comparison 
of the SFHs with the Bologna and the Baltimore solutions, which differ somewhat from each other prior to 1 Gyr ago, but still within the uncertainties.

Given that the RGB tip is so close to the completeness limit, we also tested the impact of the distance modulus on the final SFH. We found that, within the distance uncertainties derived by \cite{alo07} (18.2$\pm$1.5 Mpc), there are no significant effects on the final SFH.

\subsubsection{Region D}

\begin{figure}
\epsscale{1}
\includegraphics[width=9cm]{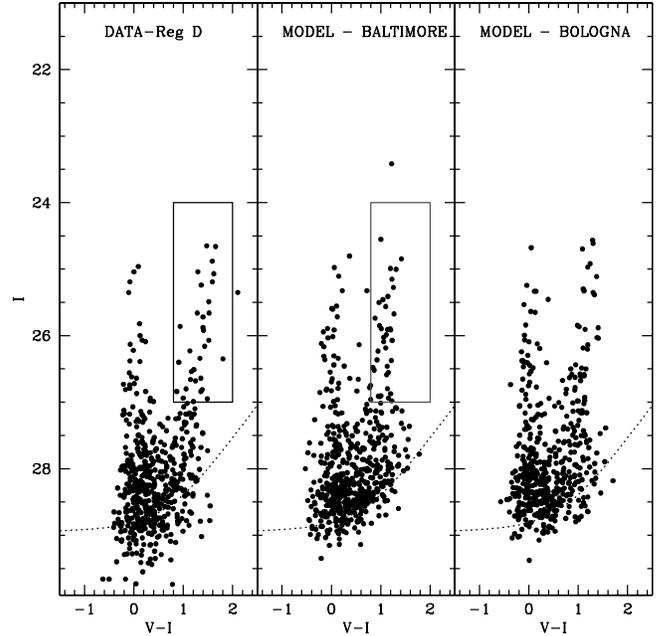}
\caption{From left to right: observed CMD for region D, best-fit synthetic CMD with the Baltimore code, best-fit CMD with the Bologna code.   
The best-fit CMDs were obtained considering only the region above 
the 20\% completeness line, indicated by the dotted curve. The box is where TP-AGB stars are located. 
\label{fig10}}
\end{figure}

\begin{figure}
\epsscale{1}
\includegraphics[width=9cm]{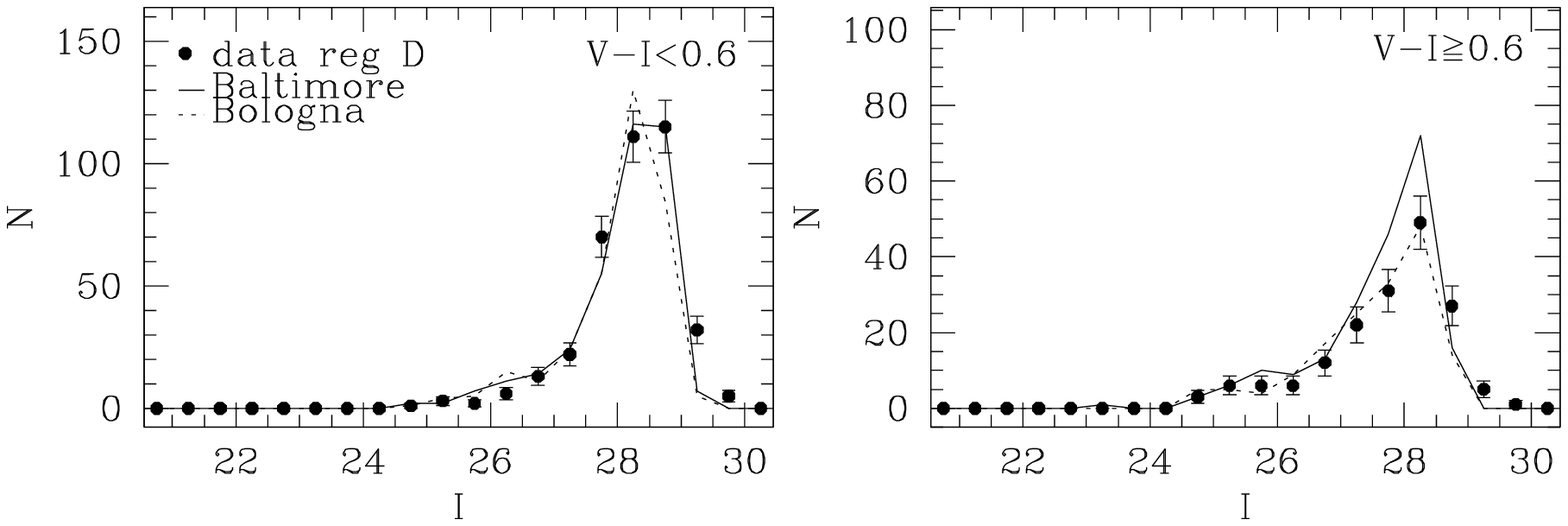}
\caption{Luminosity functions in the blue (left) and in the red (right) for the observed CMD in region D (points) and the best-fit CMDs 
obtained with the Baltimore (solid line) and the Bologna (dotted line) odes.     
\label{fig10b}}
\end{figure}

The CMD of region D in the secondary body appears quite similar to that of region C in the main body (Fig.~\ref{fig2}).
The photometry is deep enough to potentially reach RGB stars sampling the SF at epochs older than $\sim$1 Gyr. 
The comparison between the observed CMD and two synthetic CMD realizations  
from the best-fit SFH with the Baltimore and the Bologna codes is shown in Fig.~\ref{fig10}.
As for region C,  the  $\chi^2$ minimization was done using the CMD down to the 20\% completeness limit, in order to reach faint stars in the RGB phase. 
When running the Baltimore code, a pixel size of 0.25 in both color and magnitude 
was adopted and the region of TP-AGB stars at $0.8<V-I<2$, $24<I<27$ was excluded from the fit. 
On the other hand, with the Bologna code we adopted a pixel size of 0.25$\times$0.25 and 
0.5$\times$0.5 in the magnitude ranges $22<I<24$ and $24<I<29$, respectively, and used all the CMD down to the 20\% 
completeness limit for the $\chi^2$ minimization. 
Both solutions reproduce quite well the overall CMD morphology.  The agreement with the LFs in Fig.~\ref{fig10b} is quite good too, with the 
only exception of the region at $27.5<I<28.5$, $V-I\ge0.6$, where the Baltimore solution tends to over predict the counts.

\begin{figure}
\epsscale{1}
\includegraphics[width=9cm]{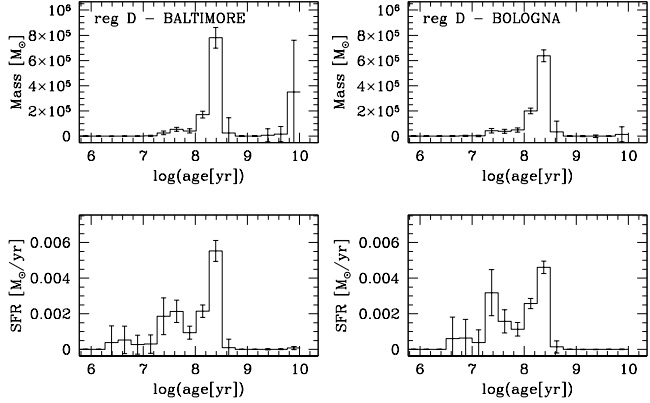}
\caption{
Best-fit SFH for region D with the Baltimore code (left panels) and with the Bologna code (right panels). From top to bottom: 
stellar mass formed and SFR in the different age bins. 
\label{fig11}}
\end{figure}

The best-fit SFHs obtained with the two different procedures are in very good agreement with each other 
(Fig.~\ref{fig11}). In both solutions, the highest peak of SF occurred $\sim$ 250 Myr ago at a rate of 
$\sim5\times10^{-3}M_{\odot}/yr$, and the SFR has been gradually decreasing since then. 
The Baltimore solution is compatible with SF during the 1$-$10 Gyr age interval, with an upper limit 
of  $\approx 10^6 M_{\odot}$ for the stellar mass (see Figure~\ref{fig11}), while the Bologna 
solution seems to exclude the presence of significant SF at these epochs. 
This implies that the fraction of the mass in old stars in region D is poorly
constrained, and might be anywhere between 0 and 40\%.

\subsubsection{Region E}

\begin{figure}
\epsscale{1}
\includegraphics[width=9cm]{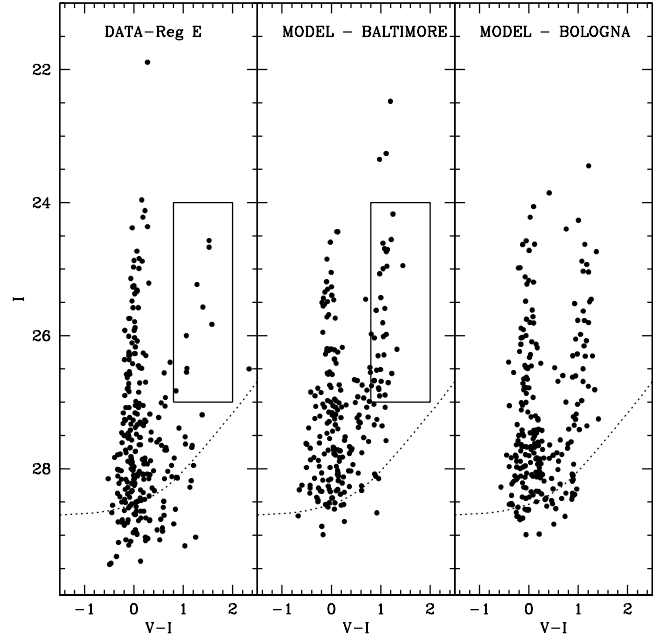}
\caption{
From left to right: observed CMD for region E, best-fit synthetic CMD with the Baltimore code, best-fit CMD with the Bologna code.   
The best-fit CMDs were obtained considering only the region above 
the 20\% completeness line, indicated by the dotted curve. The lower box is where TP-AGB stars are located. 
\label{fig12}}
\end{figure}

The comparison between the observed CMD in the most crowded region E of the secondary body  
and two synthetic CMD realizations  
from the best-fit SFH with the Baltimore and the Bologna codes is shown in Fig.~\ref{fig12}.
Due to the higher crowding than in region D, it is more difficult here to sample RGB stars.
In fact, the completeness is as low as $\sim$30\% at the RGB tip.  
The fit was done considering the CMD down to the  20\% completeness limit.
When running the Baltimore code, we excluded the region of the TP-AGB stars, and adopted a pixel size of 0.25 in both 
color and magnitude.  
On the other hand, when running the Bologna code, we considered all the CMD down to the 20\% completeness limit and 
adopted a pixel size of 0.5$\times$0.5 and 0.25$\times$0.25 in the magnitude ranges $I<26$ mag and $I\ge26$ mag, respectively.

\begin{figure}
\epsscale{1}
\includegraphics[width=9cm]{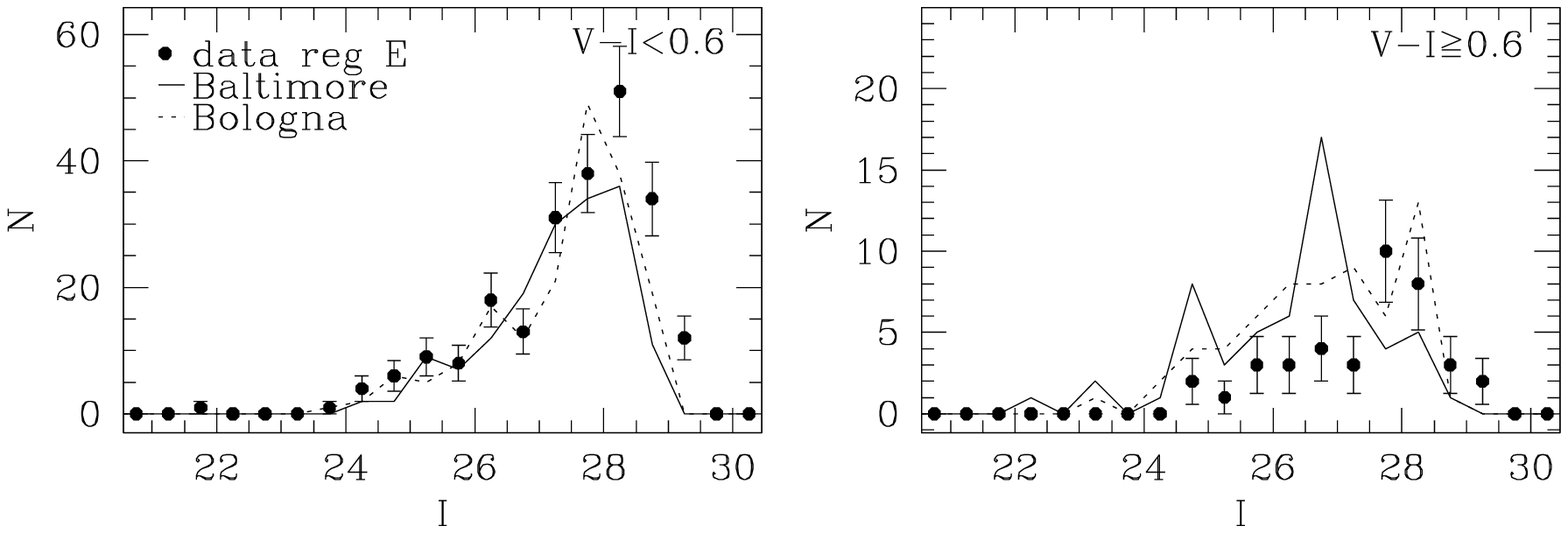}
\caption{Luminosity functions in the blue (left) and in the red (right) for the observed CMD in region E (points) and the best-fit CMDs 
obtained with the Baltimore (solid line) and the Bologna (dotted line) odes.     
\label{fig12b}}
\end{figure}

\begin{figure}
\epsscale{1}
\includegraphics[width=9cm]{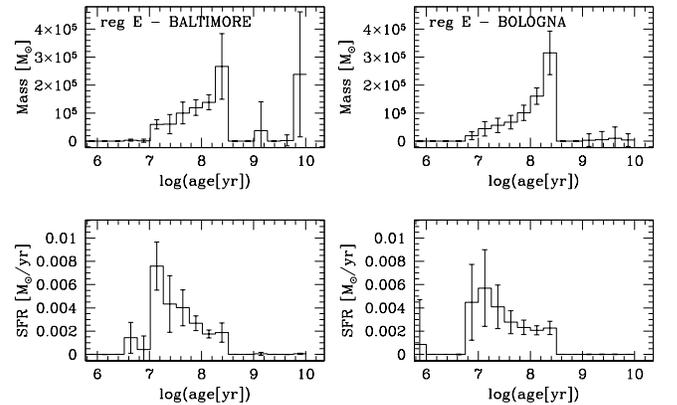}
\caption{
Best-fit SFH for region E with the Baltimore code (left panels) and with the Bologna code (right panels). From top to bottom: 
stellar mass formed and SFR in the different age bins. 
\label{fig13}}
\end{figure}

From Figs.~\ref{fig12}  and ~\ref{fig12b} we notice that the Baltimore solution tends to under predict the blue ($V-I<0.6$) counts at $I>28$ mag 
while overproducing the red ($V-I\ge0.6$) counts at  $24.5<I<27.5$; since this is the region of TP-AGB stars, that was excluded from the fit, 
the poor agreement between simulations and observations should not surprise. On the other hand, the Bologna solution seems to better reproduce 
both the overall CMD morphology and the LFs. Despite these differences, the best-fit SFHs obtained with the two procedures (Fig.~\ref{fig13}) are 
in very good agreement with each other. The highest peak of SF occurred 30-10 Myr ago at a rate of 
$\approx 7 \times 10^{-3}M_{\odot}/yr$. This is much lower than what found for the most active regions of the main body. The Baltimore solution
is compatible with the presence of SF at ages older than 1 Gyr (see Tables~\ref{tab_sfr} and ~\ref{tab_mass}).

\begin{deluxetable}{lccccc}
\tablecolumns{6}
\tablewidth{0pc}
\tablecaption{Average SFR (in $10^{-3} M_{\odot}$/yr)  in I~Zw~18 at different epochs}
\tablehead{
\colhead{Region\tablenotemark{a}} &  \colhead{Area} & \colhead{$<$10}  & \colhead{10-100} &   \colhead{0.1-1} & \colhead{1-10} \\
\colhead{} &  \colhead{[kpc$^2]$} & \colhead{[Myr]}  & \colhead{[Myr]} &   \colhead{[Gyr]} & \colhead{[Gyr]} \\
}
\startdata
A & 0.06 & 1107.0                & $>$24.7       & -                            &  -                    \\   
   &         & [1026.0]               & $>$[14.9]    & -                            &  -                    \\   

B & 0.66 & 20.0                        & 12.7         & $>$0.7          &  -                      \\ 
    &         & [23.0]                        & [16.2]          & $>$[2.5]           &  -                      \\ 

C & 0.76 & 2.0                           & 1.5             & 0.8                  & 0.2       \\
   &          & [2.0]                            & [2.1]             & [1.4]                  & [0.1]       \\

\hline
MB & 1.48 &1129.0    & $>$38.9 & $>$1.5              & $>$0.2   \\
       & &[1051.0]  & $>$[33.2] & $>$[3.9]              & $>$[0.1]   \\

\hline
D & 0.17 &          0.3  & 1.3      & 1.1  & 0.04  \\
    &      &          [0.4]  & [1.4]   & [1.0]  & [$<$0.01]  \\

E & 0.70  &         0.5  & 3.8   & 0.4  &0.03  \\
   &          &        [2.0]  & [3.0]   & [0.5]  &[$<$0.01]  \\
\hline 
SB & 0.87 & 0.8  & 5.1 & 1.5   &0.07    \\
     & & [2.4]  & [4.4] & [1.5]   &[$<$0.02]    \\
\enddata
\tablenotetext{a}{Region: MB$=$main body; SB$=$secondary body.}
\tablecomments{Values are given for the Baltimore solution and, in brackets,  for the Bologna solution.
The symbol $>$ is for lower limits.} 
\label{tab_sfr}
\end{deluxetable}

\begin{deluxetable}{cccccc}
\tablecolumns{6}
\tablewidth{0pc}
\tablecaption{Stellar mass (in $10^6 M_{\odot}$) formed in I~Zw~18 at different epochs}
\tablehead{
\colhead{Region\tablenotemark{a}} &  \colhead{$<$10}  & \colhead{10-100} &   \colhead{0.1-1} & \colhead{$>$1} & \colhead{Total} \\
\colhead{} &  \colhead{[Myr]}  & \colhead{[Myr]} &   \colhead{[Gyr]} & \colhead{[Gyr]} & \colhead{} \\
}
\startdata
A & 11.07   & $>$2.22  & -                      &  -                    & $>$13.29  \\   
   & [10.26] & $>$[1.34] & -                     & -                     &   $>$[11.60] \\ 
B & 0.20        & 1.14         & $>$0.61   &  -                     & $>$1.95     \\ 
   & [0.23]       & [1.46]       & $>$[1.68] &  -                    & $>$[3.37]     \\ 
C & 0.02                & 0.14        & 0.70         & 1.45       & 2.31  \\
    &       [0.02]       & [0.19]       & [1.25]       & [0.87]     & [2.33]  \\
 \hline
MB           & 11.29     & $>$3.50   & $>$1.31   & $>$1.45   & $>$17.55\\
                & [10.51]    & $>$[2.99] & $>$[2.93] & $>$[0.87] & $>$[17.30]\\
\hline
D & 0.003    & 0.12   & 0.98     & 0.37    & 1.47  \\
    & [0.004]  & [0.13]   & [0.87]  & [0.01] & [1.01]  \\
E &  0.005   & 0.34    & 0.40     & 0.28 & 1.02 \\
   &  [0.020]  & [0.27]  & [0.48]   & [0.02] & [0.79] \\
\hline 
SB & 0.008  & 0.46   & 1.38 & 0.65    & 2.50 \\
     & [0.024] & [0.40] & [1.35] & [0.03] & [1.80] \\
\enddata
\tablenotetext{a}{Region: MB$=$main body; SB$=$secondary body.}
\tablecomments{Values are given for the Baltimore solution and, in brackets,  for the Bologna solution. The symbol $>$ is for lower limits.} 
\label{tab_mass}
\end{deluxetable}

\section{Discussion}

In previous papers \citep{alo07,fiore10,contr11} we have tried to qualitatively characterize the different stellar populations in I~Zw~18 
and have demonstrated the presence of RGB stars, excluding 
the possibility that this is a truly primordial system in the present-day Universe. In this paper we use HST/ACS data 
in combination with the method of synthetic CMDs to quantitatively derive the SFH in I~Zw~18. Previous similar studies  
have been presented by \cite{aloisi99} and by \cite{jamet10}. 
However, the SFH derived by \cite{aloisi99} was based on shallower WFPC2 data and on the 
assumption of a distance of 10 Mpc, much closer than what is suggested by the deeper ACS CMD and by the Cepheid study. 
On the other hand, \cite{jamet10} used ACS data (only the Thuan dataset) for the derivation of the SFH in the secondary body of I~Zw~18, 
but inferred  a distance of $\sim$27 Mpc from the CMD analysis, incompatible with the distance of $\sim$18 Mpc obtained from the Cepheid light curves. In this paper we present for the first time the SFH of I~Zw~18 based on a robust distance assumption. 

In order to account for the different crowding conditions in I~Zw~18, we have selected three regions (A, B, and C) and two regions (E and D) 
in the main and secondary bodies, respectively. In the main body, region A is the most crowded one and corresponds to what has been referred to as the NW component in previous papers \citep[see e.g. Fig.~1 in ][]{papa12}, while region C is the least crowded one and is the most suitable to study the old stellar 
population. In the secondary body, region E follows the comma-shaped distribution of the brightest blue stars, while region D comprises the redder 
less crowded ``halo''.  

The first important result of this paper is an estimate of the stellar mass locked-up in old (age$>$1 Gyr) stars.  Our study indicates a mass of $\approx10^{6} M_{\odot}$ in old stars in region C, although with large uncertainties. This is about half of the total stellar mass in region C.  
On the other hand, the severe crowding in regions A and B prevents to reach RGB stars and to derive the SFH at ages older than 1 Gyr.
From this result we can derive a lower limit for the total mass in old stars in the whole main body. 
Assuming in region A and B the same surface density of old stars than in region C, and using the areas reported in Table~\ref{tab_sfr}, we obtain a mass of $\approx 2 \times 10^{6} M_{\odot}$. Since stars tend to be more concentrated toward the center, it is likely that the true mass in old stars in the main body is significantly higher than this. 
As for the secondary body, the amount of old stars is less constrained and we don't attempt to quantify it.

The second important result of our study concerns the very recent SF in I~Zw~18. The main body has been forming stars very actively in recent epochs, with an average SFR  over the last 10 Myr as high as $\approx 1 M_{\odot}/yr$ in the NW component (region A). This corresponds to a specific rate of 
$\approx2 \times10^{-5} M_{\odot} yr^{-1} pc^{-2}$, which is just a few factors lower than the rate derived in the strong star forming region 30 Doradus in the LMC. We recall however that this result relies on the assumption that all the brightest stars in the CMD are in fact individual 
objects rather than blends of two or more stars or unresolved star clusters, although we have demonstrated that the colors of the reddest ($V-I\gsim1$) bright objects are incompatible with the cluster hypothesis \citep{contr11}. Indeed, our value can be compared with the rate inferred from the 
H$\alpha$ luminosity. From \cite{cannon02} and assuming a distance of 18 Mpc, we derive a H$\alpha$ luminosity of $\sim5.9\times10^{39}$ erg/s for the NW component. Notice that this only includes the ionized gas around the peak of the continuum emission ($\approx$region A) 
but not the ionized gas extending outside the stellar component. Using the relation from \cite{kenni94} for a Salpeter's IMF in the 0.1-100 $M_{\odot}$ mass 
interval, we obtain a SFR of $\sim$0.05 $M_{\odot}/yr$, significantly lower than what was derived from the synthetic CMDs. 
We notice that the occurrence of star formation at very recent epochs in the main body is in agreement with the presence of ultra long period variables \citep{fiore10,marco10}.
A potential rate of $\approx 1 M_{\odot}/yr$ in region A implies a stellar mass of $\approx10^7 M_{\odot}$. 
This high concentration of mass is interesting in view of the results by \cite{lelli12}, who analyzed archival VLA data and found that 
the HI associated to the starburst region forms a compact fast-rotating disk. The rotation curve is flat with a steep rise in the inner parts, indicating the presence of a strong central concentration of mass. 

In the main body, the average SFR over the last 10 Myr decreases from the most crowded region A to the least crowded region C down to 
$\approx 2 \times 10^{-3} M_{\odot}/yr$. However, this is still comparable to the average rate inferred in region C over the last $\sim$ 1 Gyr.   
On the other hand, the secondary body was much less active than the main body during the last $\sim$10 Myr, in agreement with the absence of nebular emission everywhere but in its central region \citep[see also][]{papa12}. The peak of SF is found to occur $\sim$15 Myr ago in the more crowded region E and $\sim250$ Myr ago in the less crowded region D.

 \section{Summary} 
 
\begin{itemize}

\item We confirm that I~Zw~18 has started forming stars earlier than $\sim$1 Gyr ago, and possibly at epochs as old as a Hubble time. 
Thus it is not a truly young galaxy at its first bursts of star formation, as argued in previous studies \citep[e.g., ][]{it04,jamet10,papa12}. 

\item In the periphery of  I~Zw~18' s main body, where crowding is low enough to potentially detect the old population, we estimate a mass of 
$\approx10^{6} M_{\odot}$ in stars  older than $\sim$1 Gyr, accounting for about half of the total stellar mass in this region. 
On the other hand, crowding is too severe to allow the detection of such a population in the more central regions. 
 Assuming there the same surface density of old stars than in the periphery,     
we obtain a lower limit of $\approx 2 \times 10^{6} M_{\odot}$ for the total mass in old stars in I~Zw~18' s main body,
The presence of a significant amount of old stars in the secondary body is more uncertain.

\item The main body has been forming stars at a very high rate in recent epochs. 
In the most crowded NW region (region A), the average rate over the last $\sim$10 Myr was as high as $\approx 1 M_{\odot}/yr$,  
corresponding to a specific rate of $\approx2 \times10^{-5} M_{\odot} yr^{-1} pc^{-2}$. However, this result relies on the assumption 
that all the brightest stars in the CMD are in fact individual objects rather than blends of two or more stars or unresolved star clusters.

\item While in the main body the peak of activity occurred during the last $\sim$10 Myr, the secondary body was much less active at these 
epochs, in agreement with the absence of significant nebular emission. In the secondary body the peak of activity is found to occur $\sim$15 Myr ago in the more crowded central region and $\sim250$ Myr ago in the periphery. 

\item The high current SFR in I~Zw~18  explains why this galaxy is so blue and has a high ionized gas content, resembling primeval galaxies in the early Universe. Detailed chemical evolution models are required to 
quantitatively check whether the SFH from the synthetic CMDs can explain the low measured element abundances, 
both in the ionized \citep[e.g.,][]{garnett97,it98} and in the neutral interstellar medium \citep{aloisi03,lde04,james13}, or if galactic winds with loss of 
metals are needed, as originally suggested by \cite{mt85,pil93,marconi94}.

\end{itemize}

\acknowledgments
F.A. has been supported by Cofis-ASI-INAF-07, ASI I009/10/0 and PRIN-INAF-2010.
G.C., M.C. and M.T. have been also partially funded with the above grants. 
M.C., G.F. and M.T. are grateful to the International Space Science Institute (ISSI)
in Bern for the warm hospitality during Team meetings.

\clearpage

\end{document}